\documentclass[bibyear]{aa}  
\usepackage{graphicx}
\usepackage{txfonts}
\usepackage{enumerate}
\usepackage{txfonts}
\usepackage{graphicx}

\begin{document}

\title{The rotation-metallicity relation for the Galactic disk as measured in the Gaia DR1 TGAS and APOGEE data}
\titlerunning{Rotation-Metallicity Relation with Gaia TGAS-APOGEE}
  
\author{Carlos Allende Prieto\inst{1,2,3}, Daisuke Kawata\inst{3} and Mark Cropper\inst{3}}
\authorrunning{Carlos Allende Prieto et al.}

   \institute{Instituto de Astrof\'{\i}sica de Canarias,
              V\'{\i}a L\'actea, 38205 La Laguna, Tenerife, Spain\\              
         \and
             Universidad de La Laguna, Departamento de Astrof\'{\i}sica, 
             38206 La Laguna, Tenerife, Spain \\
         \and 
             Mullard Space Science Laboratory, University College London, 
             Holmbury St. Mary, Dorking, Surrey, RH5 6NT, UK
             }

%

   \date{submitted September 26, 2016; accepted November 2, 2016}

 
  \abstract
   {}
   {Previous studies have found that the Galactic rotation velocity-metallicity (V-[Fe/H]) relations 
   for the thin and thick disk populations show  
negative and positive slopes, respectively. 
The first Gaia Data Release includes the Tycho-Gaia Astrometric Solution (TGAS) information, 
which we use to analyze the V-[Fe/H] relation for a strictly selected sample with high enough astrometric accuracy. 
We aim to present an explanation for the slopes of the V-[Fe/H]  relationship}  
   {We have identified a sample of stars with accurate  Gaia TGAS data  
and SDSS APOGEE [$\alpha$/Fe] and [Fe/H] measurements. 
We measured the V-[Fe/H] relation for  thin and thick disk stars classified on the basis of 
their [$\alpha$/Fe] and [Fe/H] abundances.}
   {We find dV/d[Fe/H]$=-18\pm 2 $~km~s$^{-1}$~dex$^{-1}$ for stars in the 
thin disk and dV/d[Fe/H]$=+23\pm 10$~km~s$^{-1}$~dex$^{-1}$ for thick disk 
stars, and thus we confirm the different signs for the slopes. 
The negative value of dV/d[Fe/H] for thick disk stars is 
consistent with  previous work, but the combination of TGAS and APOGEE 
data provide higher precision, even though systematic errors could exceed 
 $\pm 5$ km s$^{-1}$ dex$^{-1}$. Our average 
measurement of dV/d[Fe/H] for local thick disk stars shows a somewhat 
flatter slope than in previous studies, but we confirm a significant
spread and a dependence of the slope on the [$\alpha$/Fe] ratio of 
the stars. Using a simple N-body model, we demonstrate that the observed 
trend for the thick and thin disk can be explained by the measured  
radial metallicity gradients and the correlation between 
orbital eccentricity and metallicity in the thick disk.
}
   {We conclude that the V-[Fe/H] relation for thin disk stars is well determined from our 
   TGAS-APOGEE sample, and a direct consequence of the radial metallicity gradient and 
   the correlation between Galactic rotation and mean Galactocentric distance.  
   Stars formed farther away from the solar circle tend to be near their orbital pericenter, 
   showing larger velocities and on average lower metallicities, while those closer to the 
   Galactic center are usually closer to their orbital apocenter, therefore moving slower 
   and with higher metallicities. 
The positive dV/d[Fe/H] for the thick disk sample is likely connected to the correlation 
between orbital eccentricity and metallicity for these stars. 
}


   \maketitle
%

\section{Introduction}
\label{Intro}

The stellar populations of the thin and thick disk of the Milky Way exhibit a
significant overlap in metallicity ([Fe/H]), age, and kinematics (e.g., Bensby et al. 2014).
The distinction between these two populations is most obvious in the combined
[Fe/H]-[$\alpha$/Fe] space, the Galactic rotation velocities ($V_{\phi}$) 
in terms of both average values and dispersion, and the vertical velocity dispersion
($V_z$). However, at high metallicity the $\alpha$-element content of thin
and thick disk stars are similar, and confusion between the two populations
can be severe. 
Furthermore, the separation between metal-poor thick disk
stars and members of the halo is not trivial since their metallicity and 
velocity distributions, despite their significant differences, overlap.

An intriguing statistical relationship has been found between the Galactic
rotation of disk stars and their metallicity (V-[Fe/H] relation). This 
relationship shows opposite
signs for the thin and the thick disk components 
(Spagna et al. 2010, Lee et al. 2011, Adibekyan et al. 2013).
Understanding the origin of this difference in the sign of the relationship is important. 
It may be spurious:  confusion between halo and metal-poor thick disk stars can 
induce a false V-[Fe/H] correlation in thick disk samples, since halo stars are statistically
more metal-poor and show essentially no Galactic rotation. Likewise, leakage of 
thin disk stars into samples of thick disk stars can create a false gradient, 
or mask an existing one, and although
statistically more unlikely in local samples, thick disk stars confused with 
thin disk members can lead to similar mistakes. 

The situation is further complicated by the fact that there is no consensus about the
regions of the [Fe/H]-[$\alpha$/Fe] space that thin and thick disk stars
 occupy, and not even about whether they occupy distinct regions. For example, 
local samples studied
by Fuhrmann (2011 and references therein) and Adibekyan et al. (2013) have led these authors 
to identify three distinct chemical groups of stars: one with lower [Fe/H] and higher 
[$\alpha$/Fe] abundances (thick disk), one with higher [Fe/H] and relatively 
lower [$\alpha$/Fe] values (thin disk), and a third with intermediate chemistry. 
On the other hand, other local samples such as 
those by Bensby et al. (2011, 2014) or Ramirez et al. (2013) appear to show exclusively 
two sequences in [Fe/H]-[$\alpha$/Fe] space, and  the results 
from non-local high-resolution studies of very large samples such as those from 
APOGEE (Majewski et al. 2016; Hayden et al. 2014, 2015; Anders et al. 2015; 
Nidever et al. 2013) and the {\it Gaia}-ESO Survey (Gilmore et al. 2012; Recio-Blanco et al. 2014) 
tend more towards this latter scenario. 
Adding to the confusion, analyses based on photometry or lower-resolution spectroscopy tend to  
portray the two disk populations as a single  one with a continuous range of correlated  
chemical and kinematical properties (Ivezi\'c et al. 2008; Bond et al.2010;  Bovy et al. 2012).

The motivation for this paper is to establish whether the signs in the V-[Fe/H] relationship 
are indeed opposite between thin and thick disks, and if so, to achieve an understanding as 
to why this may be the case.
We re-examine the separation in chemistry and kinematics between 
the two disks taken advantage of recently published near-infrared ($H$-band) 
spectroscopy from APOGEE for nearby stars with astrometric data in  
 the combined {\it Tycho}-{\it Gaia} astrometric solution (TGAS; Michalik, Lindegren, \& Hobbs 2015; 
 Gaia Collaboration, Prusti et al. 2016).
The TGAS data provide Hipparcos-quality astrometry, with uncertainties in parallaxes 
typically under 1 milliarcsecond, and those in proper motions under 1 milliarcsecond 
per year, for a sample more than ten times larger and three magnitudes fainter than Hipparcos. 
The APOGEE observations typically focus on fainter stars than those in the TGAS catalog 
but there is a significant overlap between the two. 

We pay particular attention to the reliability of previously identified correlations
between the Galactic rotational velocities of stars and their metallicity. Using ad hoc
N-body numerical simulations of Milky-Way like disks, we discuss
what may be driving the observed patterns. 
The paper is structured as follows. Section \ref{data} describes the catalogs of
observations we employ. Section \ref{analysis} examines the thin and thick disk separation
of the sample stars and our analysis. Section \ref{model} describes our models and their predictions, 
while Section \ref{summary} summarizes our findings and conclusions.

\section{Observational data}
\label{data}

Our analysis is based on two types of observations, astrometry from 
TGAS and spectroscopy from APOGEE. More details for  each of these data sources are given below.

\subsection{Astrometry}
\label{tgas}

The Hipparcos satellite was launched in 1989 and the results were published as the 
Hipparcos and Tycho catalogs by ESA (1997). The mission provided astrometric solutions for 
more than 1$\times 10^5$ stars down to 
12th magnitude (complete to about 8th magnitude). The Hipparcos observations were 
later reprocessed to produce an improved catalog after improvements by van Leeuwen 
and colleagues (van Leeuwen 2007a,b) in modeling
the 
satellite's attitude. 
Observations from the Hipparcos starmapper led to the creation of the original
Tycho catalog, which was later enhanced, making the Tycho-2 catalog (H\o{}g et al. 2000a,b).
Tycho-2 includes $2.5\times10^6$ stars, and their photometry in the $B_T$ and $V_T$ bands. 
It is essentially (99\%) complete down
to 11th magnitude.

The combination of the Tycho-2 catalog with {\it Gaia} observations (Gaia Collaboration, 
Prusti et al. 2016) obtained
over the first year of the mission have led to the TGAS (Michalik et al. 2015; 
Gaia Collaboration, Brown et al. 2016; Lindegren et al. 2016) as mentioned in 
Section \ref{Intro}. This catalog includes  astrometric
parameters with absolute random uncertainties similar to, or better than,
those in the Hipparcos 
catalog, but for the fainter stars in Tycho-2. The median statistical uncertainties
in parallaxes and proper motions are approximately 0.3 mas and 1 mas yr$^{-1}$, respectively, 
with an additional systematic uncertainty of about 0.3 mas in the parallaxes.

\subsection{Spectroscopy}
\label{apogee}

The Apache Point Observatory Galactic Evolution Experiment 
(Majewski et al. 2016) started
in 2011 as part of the Sloan Digital Sky Survey (SDSS-III; Eisenstein et al. 2011). 
The project
makes use of a multi-object (300-fiber) high-resolution near-infrared spectrograph 
to gather stellar spectra. The observations from the first three years of operation 
have been published as the SDSS-III Data Release 12 (Alam et al. 2015; Holtzman et al. 2015) 
and the SDSS-IV Data Release 13 (SDSS Collaboration 2016; Holtzman et al. 2016). 

In this work we have focused on using the APOGEE DR12 overall metallicity and 
$\alpha$-element abundance, 
since these have already being thoroughly tested, and reference literature 
studies on 
Galactic abundance gradients are available for them (Hayden et al. 2014, 2015; 
Anders et al. 2014). 
Nevertheless, we have checked how much our results would change if we adopted the latest 
data release (DR13;  SDSS Collaboration. 2016).

The APOGEE Stellar Parameters and Chemical Abundances Pipeline 
(ASPCAP; Garc\'{\i}a P\'erez et  al. 2016) derives the most relevant stellar 
parameters simultaneously 
(including the overall metallicity, carbon, nitrogen, and overall 
$\alpha$-element abundances), proceeding in a second step to derive abundances for
other elements, rederiving those for carbon, nitrogen and individual $\alpha$ elements. 
The typical statistical uncertainties in the APOGEE metallicities and 
$\alpha$-to-iron ratios are approximately 0.01 dex for stars with metallicities 
in the range $-0.6<$[Fe/H]$<0.0$ and effective temperatures $4000 < T_{\rm eff}<4300$ K, 
increasing to about 0.05 dex at $T_{\rm eff}\sim 4800$ K (Holtzman et al. 2015; 
Bertran de Lis et al. 2016). Systematic errors could reach 0.1-0.2 but trends with 
effective temperature have been largely removed using observations of open cluster
member stars.

The use of particular  elements such as iron or oxygen allows a cleaner comparison 
with  chemical evolution models incorporating detailed supernova yields. 
We have found, however, that using the overall metallicity  and $\alpha$-element 
abundances derived in the first  ASPCAP step can provide higher precision for 
separating stellar populations from their compositions. For example, 
we find in general a tighter separation between thin and thick disk stars in 
the APOGEE "$\alpha$" than in individual $\alpha$ elements such as oxygen or 
magnesium. This is not surprising, since this average $\alpha$ abundance
combines information from more lines than any single $\alpha$ element. 
In this paper we will therefore use the overall metallicity and 
$\alpha$ enrichment derived in the first ASPCAP step, and for simplicity 
we refer  to them as [Fe/H] and [$\alpha$/Fe].

\begin{figure}[t!]
\centering
{\includegraphics[width=8cm]{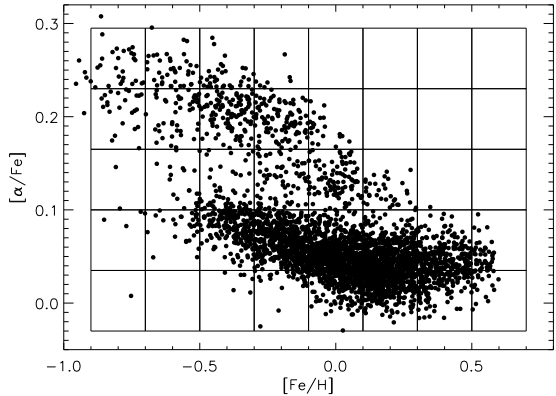}}
\caption{Distribution of overall [Fe/H] and [$\alpha$/Fe] for the 
APOGEE-TGAS sample (limited to $\sigma(p)/p<0.3$) described in Section \ref{analysis}.}
\label{box}
\end{figure}

\section{Analysis}
\label{analysis}

We study the correlation between kinematics and chemistry for stars in common 
between TGAS and APOGEE.  We crossed the APOGEE DR12 catalog with  
TGAS, finding 21,186  sources in common. 
We retained the giant stars in the sample, for which the APOGEE abundance measurements
are more reliable and whose parameters have been carefully calibrated taking advantage of 
Kepler asteroseismic information (Pinsonneault et al. 2015) and other reference data. 
The bulk of the dwarf stars that are both in APOGEE and TGAS are warm F and A-type stars,  
which are not very useful for chemical analysis in the H-band. 

We selected stars with surface gravity of $\log g< 3.8$, 
effective temperature of $T_{\rm eff} < 5500$ K, 
and relative uncertainties in the parallaxes smaller than 30\%. We adopt the 
uncertainties for the parallaxes given in the TGAS {\it Gaia} DR1 catalog, added 
in quadrature with a systematic uncertainty of 0.3 mas as recommended.
We further limited the sample 
to stars with metallicities [Fe/H]$>-1$, to focus on the disk population 
avoiding interlopers from the halo.  The final sample includes 3621 stars.

\begin{figure*}[t!]
\centering
{\includegraphics[width=10cm,angle=90]{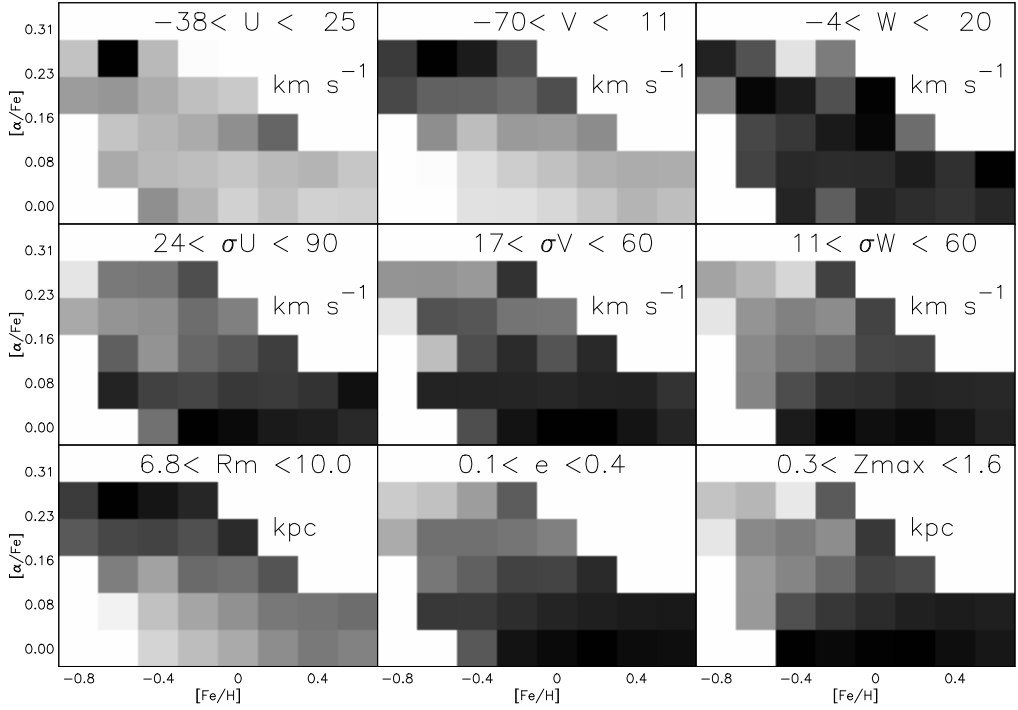}}
\caption{Kinematics for the APOGEE-TGAS stars. The top
panels show the average values in the boxes introduced in Fig. \ref{box} for the 
Galactic velocity components UVW. The middle panels show the dispersion in the same
velocities. The bottom panel shows the orbital elements: the guiding center (Rm), eccentricity (e), 
maximum distance from the Milky way plane (Z$_{\rm max}$)). The grayscale is linear and 
spans the numerical ranges indicated in each panel from the minimum (black) to the maximum 
(white -- just above the range of the data).}
\label{kin} 
\end{figure*}

Fig. \ref{box} shows the distribution of the sample in the [$\alpha$/Fe]-[Fe/H]
plane, and as in other samples, stars may be divided into two main populations, one 
with lower $\alpha$/Fe abundance ratios and another with higher ratios, associated
with the thin and thick components of the disk, respectively. 
As discussed in Section~\ref{Intro}, some have argued that there is a distinct
third group in-between these two, corresponding to stars with intermediate $\alpha$/Fe 
abundance ratios, here evident at [Fe/H] $\sim $ 0 and [$\alpha$/Fe] 
$\sim 0.12$. Haywood (2013), Haywood et al. (2016) and Feuillet et al. (2016) show 
that that group of
stars indeed shows intermediate ages between those in the lower and higher [$\alpha$/Fe]
sequences. However, that does not resolve the issue of whether they belong to 
a distinct third population. 

To examine whether those stars should be associated with those with higher $\alpha$/Fe 
ratios or not,  we consider the stellar kinematics. We combined the TGAS 
astrometry and the APOGEE  radial velocities to calculate the Galactic velocities of 
the stars with respect to the local standard of rest (LSR). We adopted 
for the solar motion
relative to the LSR the values of Sch\"onrich et al. (2010), 
(U,V,W)$_{\odot} = (11.1, 12.24, 7.25)$, and followed the 
recipe described by Johnson \& Soderblom (1987). The impact of the uncertainties 
in the astrometry on the derived Galactic velocities is modest, with mean 
(and standard deviation) of the uncertainties in V of 2 (2), 4 (4) and 6 (5) km s$^{-1}$ 
for sub-samples defined according to the relative uncertainties in the 
parallax $\sigma(p)/p =$ 0.1, 0.2 and 0.3, respectively, which is significantly 
less than the velocity spread of the disk.

The top panel of Fig. \ref{kin}
shows the average radial (U), azimuthal (V), and vertical (W) velocity components 
for each of the boxes shown in Fig. \ref{box}, while the middle panels show
the corresponding dispersions. A box size of 0.2 dex in [Fe/H] and 0.065 dex in 
[$\alpha$/Fe] has been chosen as a trade-off between adequate statistical errors and our 
ability to sample variations in kinematics as a function of chemistry. It also gives 
us two average values for each of the main high-$\alpha$ and low-$\alpha$ populations at
any given [Fe/H], which provides a consistency check. 
As anticipated from the discussion in Section~\ref{Intro}, 
the most relevant quantities that allow a separation between the thin (low-$\alpha$) and 
thick disk (high-$\alpha$) components are the Galactic rotation (V), and the dispersion 
in all components ($\sigma$U, $\sigma$V and $\sigma$W). 
Nevertheless, to the level we can say 
with these data, the boxes with intermediate
values of [$\alpha$/Fe] tend to have intermediate kinematics. 
There is even
some indication that the stars with intermediate $\alpha$-to-iron ratios are 
kinematically closer to the thin disk than to the thick disk population.

We infer orbital parameters for all our chosen sample
using our derived space velocities and {\tt galpy}\footnote{http://github.com/jobovy/galpy} (Bovy 2015). The resulting
mean orbital radius (R$_{\rm m}$), eccentricities (e), and maximum height from the Galactic plane (Z$_{\rm max}$))
are shown in the bottom panel of Fig. \ref{kin}. The average values for R$_{\rm m}$ and Z$_{\rm max}$) 
of the stars in the box at [Fe/H] $\sim $ 0 and [$\alpha$/Fe] $\sim 0.12$ have again 
intermediate values. 
The mixed values for the kinematics and orbital parameters of the stars
with intermediate values of [alpha/Fe] are not due to confusion between the
thin and thick disk populations, since the APOGEE statistical errors  in [Fe/H] and [$\alpha$/Fe] are about 0.01-0.03 dex.

\begin{figure}[t!]
\centering
{\includegraphics[width=8cm]{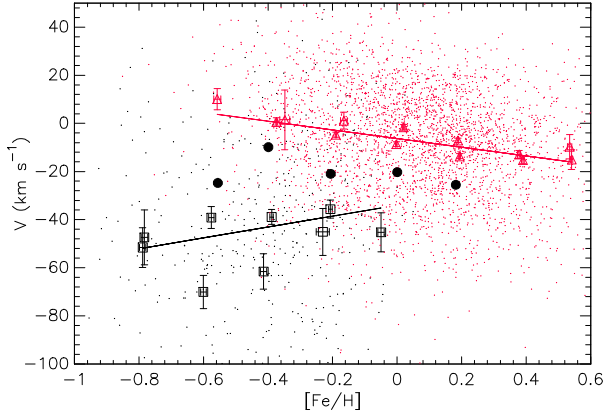}
}
\caption{Galactic rotation velocity derived for the APOGEE-TGAS (DR12) sample. 
Stars chemically associated with the thin and thick disks are 
shown as black and red dots, respectively. The average values found in the boxes 
introduced in Fig. \ref{box} are shown as triangles, and the straight lines correspond to 
linear least-squares fits to the average values. The filled circles correspond 
to the intermediate-[$\alpha$/Fe] stars not included in the thin or thick
disk samples. The average values typically come in pairs at any given [Fe/H], 
since both the thin and thick disk populations are covered by two rows of boxes in Fig. 1.}
\label{vphi}
\end{figure}

We conclude that, at least with the APOGEE-TGAS data set, it is hard to decide whether 
the intermediate $\alpha$ stars are part of the thick disk or not. We conservatively 
take the approach of adopting as thin and thick disk stars only those with extreme 
[$\alpha$/Fe] values: thin disk are taken as stars with [$\alpha$/Fe]$<0.1$ and 
thick disk stars as those with $+0.17<$[$\alpha$/Fe]$<+0.3$.
We measure the mean velocity and velocity dispersion for each population 
(see Table 1), and  make use of the mean Galactic rotation velocities measured in 
each box to examine their dependence on stellar 
metallicity. This is illustrated in Fig. \ref{vphi}, where different symbols are used
for individual stars, and mean values for the boxes.
Similarly to Lee et al. (2011), Recio-Blanco et al. (2015), or Adibekyan et al. (2013), 
linear trends are apparent after binning. However, the scatter among individual stars 
in a given population is quite large, in particular for the thick disk. 

The binned data are obtained by computing average values and their uncertainties, 
assuming a normal distribution ($\sigma/\sqrt{N}$), both axes. These data 
are fit with a linear relationship taken the uncertainties in both axes into account. 
The linear fittings indicate 
dV/d[Fe/H]$=-18 \pm 2 $ km s$^{-1}$ dex$^{-1}$ for the thin disk, and 
dV/d[Fe/H]$=+23 \pm 10 $ km s$^{-1}$ dex$^{-1}$ for the thick disk. The value for the thin
disk is of a very high significance, and in excellent agreement with previous results
in the literature, e.g., $-17 \pm 4$ km s$^{-1}$ dex$^{-1}$ by Adibekyan et al. (2013), 
or $-17\pm 6$ by Recio-Blanco et al. (2015), but slightly discrepant with $-22 \pm 3$ by Lee et al. (2011). 
The result for the thick disk 
stars is more uncertain but points to a much weaker variation of the rotation V-[Fe/H] relation
than those reported in previous studies, and is discussed below.

The average values for the stars that fall in boxes with intermediate 
$0.1 \le $[$\alpha$/Fe]$\le 0.17$ ratios are shown with filled circles in Fig. \ref{vphi}.
In addition to intermediate values for the $\alpha$/Fe ratios, and intermediate metallicity
and age distributions, these stars exhibit intermediate values for the Galactic rotation 
velocities and their dependence on [Fe/H].

\begin{table}
\centering
\caption{Mean and standard deviation for the TGAS-APOGEE sample with uncertainties in 
parallax smaller than 30\%. Uncertainties are derived from statistics, 
and the variation found when changing the limit in the uncertainty of the parallaxes.}
\begin{tabular}{lccc}
\hline
\hline
                   &  U        &   V      &     W        \\
                   & \multicolumn{3}{r}{(km s$^{-1}$)} \\
\hline
thin disk average     &  $10\pm 2$   &   $-8 \pm 1$  &  $0 \pm 1$ \\
thin disk std. dev.  &  $37 \pm 2$  &   $23\pm 1$   &  $18\pm 1$ \\
thick disk average    &  $2\pm 3$   &  $-45\pm 4$   &  $3 \pm 3$ \\
thick disk std. dev. &  $62\pm 4$  &  $39 \pm 1$  &   $40 \pm 1$ \\
\hline
\end{tabular}
\end{table}

The Hipparcos catalog, despite its small size, offers independent support for the 
results derived for the TGAS stars. There are 654 giant stars in common between 
Hipparcos and APOGEE -- basically the sample described by 
Feuillet et al. (2016), obtained for validation of the results from the APOGEE pipeline. 
Adopting exactly the same criteria described for the TGAS-APOGEE sample, we arrive at 
dV/d[Fe/H]$=-10 \pm 4 $ km s$^{-1}$ dex$^{-1}$ for the thin disk, and 
dV/d[Fe/H]$=+44 \pm 44 $ km s$^{-1}$ dex$^{-1}$ for the thick disk. The slope
for the thick disk is significantly more uncertain than the 
the TGAS-APOGEE result, but consistent with them within the uncertainties. 
The thick disk gradient is shallower than in the TGAS-APOGEE sample.   

The most recent data release of the SDSS, DR13 (Albareti et al. 2016), made public on July 2016, 
contains the same APOGEE stellar sample as DR12, but upgrades to the data reduction and 
analysis pipelines. If we replace the DR12
abundances (which in the context of this paper are limited to [Fe/H] and [$\alpha$/Fe])
by those in DR13 and repeat the analysis for the $\sigma(p)/p<0.3$ TGAS sample we arrive at 
dV/d[Fe/H]$=-23 \pm 2 $ km s$^{-1}$ dex$^{-1}$ for the thin disk, and 
dV/d[Fe/H]$=+33 \pm 15 $ km s$^{-1}$ dex$^{-1}$ for the thick disk, which are
statistically consistent with the results for DR12, but show that systematic
erros associated with the metallicity scale can easily amount to $\sim 5$ 
km s$^{-1}$ dex$^{-1}$. 

The uncertainties in the astrometry have a modest effect on the derived kinematics, 
and we have chosen to retain stars with relative uncertainties in the parallaxes better than 30\%. 
Since the uncertainties in proper motion are tightly correlated with those in parallax, and both 
are tied to the stellar brightness, a simple limit in the parallax uncertainty involves a more 
general limit on the overall astrometric quality.  
If we were to enforce a more strict limit on the uncertainties, say 10\% in relative parallax,
the sample would be severely reduced to 547 stars. This will still provide a statistically 
robust result for the thin disk of dV/d[Fe/H]$=-18 \pm 5 $ km s$^{-1}$ dex$^{-1}$, but an 
inconclusive slope for the thick disk stars: dV/d[Fe/H]$=+21 \pm 45 $ km s$^{-1}$ dex$^{-1}$. 
Similarly, retaining stars with relative parallax uncertainties under 20\% will lead to 
dV/d[Fe/H]$=-18 \pm 2 $ km s$^{-1}$ dex$^{-1}$ and 
dV/d[Fe/H]$=+11 \pm 13 $ km s$^{-1}$ dex$^{-1}$ for the thin and thick disks,
respectively. However, if we relax the limits on the astrometric quality
to enforce only a 40\%, 50\% or 100\% parallax error, progressively increasing the sample size,  
the results remain robust, as reflected in Table 2.
This weak dependence of our results on the astrometric uncertainties is 
the result of the increase in sample size associated with imposing more relaxed
limits, and the fact that a large fraction of the sample concentrates 
towards $l=90$ and $b=0$ -- about 1/3 of the stars are within 30 degrees from that direction-- 
mainly due to the extensive APOGEE program to follow-up Kepler stars
(APOKASC; see Pinsonneault et al. 2015). 

It is apparent that the average rotation velocity for stars in the 
boxes centered 
at [Fe/H]=$-0.6$ and $-0.4$ (squares at those  values in Fig. \ref{vphi}) 
may split in two groups, one with high and one with low V velocity. 
We observe that the larger V velocities correspond to the boxes with the highest 
[$\alpha$/Fe] values (those centered at [$\alpha$/Fe]=$+0.26$). This
is in line with the findings by Recio-Blanco et al. (2015), who found a steeper positive 
value for dV/d[Fe/H] for thick disk stars with lower [$\alpha$/Fe] ratios.

\begin{table}
\centering
\caption{Mean dV/d[Fe/H] values derived for the thin disk ([$\alpha$/Fe]$<0.1$) 
and thick disk ([$\alpha$/Fe]$>0.17$) stars for subsamples defined by an upper 
limit to the relative uncertainty in the relative {\it Gaia} TGAS parallaxes.}
\begin{tabular}{lrrrr}
\hline
\hline
           & \multicolumn{2}{c}{Thick disk} &  \multicolumn{2}{c}{Thin disk}            \\
   e(p)/p  &  dV/d[Fe/H]  & N & dV/d[Fe/H]  & N \\
\hline
    0.1    &  $+21\pm45$            & 35  & $-18\pm5$ &  473   \\
    0.2    &  $+11\pm13$               & 190  & $-18\pm2$ & 1575    \\
    0.3    &  $+23\pm10$           & 401  & $-18\pm2$ & 2950    \\
    0.4    &  $+23\pm8$               & 621  & $-18\pm2$ &  3984   \\
    0.5    &  $+30\pm7$               &  779  & $-19\pm1$ & 4658    \\
    1.0    &  $+32\pm7$               & 1053   & $-17\pm1$ &  5688   \\
\hline               
\end{tabular}
\end{table}

Biases may arise because stars with intermediate 
[$\alpha$/Fe]
abundance ratios show intermediate kinematics between the high-$\alpha$ ([$\alpha$/Fe]$>0.18$)
and the low-$\alpha$ populations. It is easy to see by inspection of Fig. \ref{vphi}
that including the stars with intermediate [$\alpha$/Fe] and relatively higher [Fe/H] will
enhance the derived   dV/d[Fe/H] gradient for the thick disk population. 
In addition, as mentioned
in Section~\ref{Intro}, halo stars are more likely to enter in the low-metallicity side 
of the thick disk distribution, which can lead to an artificially steeper slope.
This issue might 
already be mildly affecting our analysis, since capping the thick disk TGAS sample
to $V>-100$ km s$^{-1}$ would slightly flatten our slope for this population from 
dV/d[Fe/H]$=+23 \pm 10$
to
dV/d[Fe/H]$=+14 \pm 9$, but we stress that such a limit in $V$ would be certainly extreme. 

Our data permit a fairly reliable determination of the correlation between orbital 
eccentricity and metallicity for each sub-sample. 
Here, eccentricity is calculated from the orbital apocenter and pericenter as
$e=$(R$_{\rm apo}$-R$_{\rm peri}$)/(R$_{\rm apo}$+ R$_{\rm peri}$).
For the thin disk stars we find
d$e$/d[Fe/H] $= -0.05 \pm 0.01$ dex$^{-1}$, while for the thick disk sample we find 
d$e$/d[Fe/H] $= -0.11 \pm 0.03$ dex$^{-1}$. These results are  
consistent with the measurements reported by Adibekyan et al. (2013), 
d$e$/d[Fe/H] $= -0.023 \pm 0.015$ dex$^{-1}$ 
for the thin disk stars and d$e$/d[Fe/H] $= -0.184 \pm 0.078$ dex$^{-1}$ for the thick disk stars.
 We note  that Adibekyan et al. include disk candidate stars down to 
[Fe/H]$\simeq -1.4$, a domain in which it becomes difficult to 
eliminate halo stars from  local samples, which could produce 
a pronounced increase in the eccentricity of the metal-poor stars in the sample.

We need d[Fe/H]/d$e$ for a convenient parameterization in 
the models described below, and  since there is significant
scatter in the [Fe/H]-$e$ relation, the best fit d[Fe/H]/d$e$ does not 
equal 1/(d$e$/d[Fe/H]). We measure d[Fe/H]/d$e=-0.34\pm0.08$ 
for the thick disk stars in the TGAS-APOGEE sample.

\section{Model comparison}
\label{model}

To understand our observed dV/d[Fe/H] trend for the thick and thin disk populations, 
we compare our results with a snapshot of an N-body simulation. The N-body simulation model used 
is the same as Model A of Kawata et al. (2016), but with a 
different initial velocity dispersion for the 
thick disk particles with $\sigma_{\rm U}^2/\sigma_{\rm W}^2=1$ instead of 
$\sigma_{\rm U}^2/\sigma_{\rm W}^2=2$
for Model A, since this leads to a more realistic ratio between the 
azimuthal and vertical velocity dispersions.
We use our  Tree N-body code, {\tt GCD+} (Kawata \& Gibson 2003; 
Kawata et al. 2013) for the N-body simulation.

We initially set up an isolated disk galaxy which consists of stellar thick and thin exponential disks, 
with no bulge component, in a static Navarro, Frenk \& White (1997) 
dark matter halo potential (Rahimi \& Kawata 2012; Grand et al. 2012). 
The initial scale length and scale heights of thick (thin) disks are set to be 
$R_{\rm d,thick}=2.5$ kpc ($R_{\rm d,thin}=4.0$ kpc) 
and $z_{\rm d,thick}=1.0$~kpc ($z_{\rm d,thin}=0.35$~kpc), 
respectively. The mass of the thick and thin disks are 
$M_{\rm d,thin}=4.5\times10^{10}$~M$_{\rm \sun}$ 
and $M_{\rm d,thick}=1.5\times10^{10}$~M$_{\sun}$. 

We used a snapshot at $t=1$~Gyr,  after spiral arms have developed, 
but before a bar forms. The model is a pure N-body simulation, 
and there is no gas component or growth of the stellar disk for simplicity. 
We chose this particular snapshot
because the azimuthal, $\sigma_{\rm V}$, and vertical, $\sigma_{\rm W}$, velocity dispersion for the thin 
and thick disk at the Solar radius is similar to that observed in the Milky Way. Unfortunately, 
we do not have an N-body simulation having all three components of the 
velocity dispersion consistent with the Milky 
Way disk. We prioritized $\sigma_{\rm V}$ and $\sigma_{\rm W}$, and compromised on $\sigma_{\rm U}$. 

We tagged the particles belonging to the thin disk component at the start of
the simulation. 
From our TGAS-APOGEE sample, we measured the radial metallicity gradient 
as a function of mean orbital Galactocentric radius, R$_{\rm m}$, 
and the vertical gradient as a function of Z$_{\rm max}$),
and found d[Fe/H]/dR$_{\rm m}=-0.053 \pm 0.004$~dex~kpc$^{-1}$ and 
d[Fe/H]/dZ$_{\rm max}$)$=-0.34 \pm 0.03$~dex~kpc$^{-1}$.
We then assigned metallicities to the particles, according to their Galactocentric distances,  
[Fe/H]$=-0.05 \times {\rm R}_{\rm m}-0.3\times {\rm |Z|}+0.4$~dex, with a dispersion of 0.2 dex, 
where R$_{\rm m}$=(R$_{\rm apo}$+R$_{\rm peri}$)/2 is the mean radius of the orbit, R$_{\rm apo}$
and R$_{\rm peri}$ are the particle's apo- and pericenter radii, respectively, 
and $|$Z$|$ is the current 
vertical height. We ran a test particle simulation for the selected particles under the gravitational  
potential calculated from the frozen particle distribution, 
and analyzed R$_{\rm apo}$ and R$_{\rm peri}$.
Note that we used the current vertical height instead of Z$_{\rm max}$) for simplicity. 

We selected particles within a ring at $7.5<$R$<8.5$~kpc and $|{\rm Z}|<0.5$~kpc to mimic 
a volume roughly consistent with that occupied by our TGAS-APOGEE stars, assuming that the Galactocentric 
radius of the Sun is 8~kpc.  The velocity dispersion of the selected sample of particles 
is $(\sigma_{\rm U}, \sigma_{\rm V}, \sigma_{\rm W})=(28, 22, 18)$ km~s$^{-1}$. As mentioned above, the radial velocity dispersion, $\sigma_{\rm U}$, is smaller than the observed $\sigma_{\rm U}$ 
of the thin disk population of the Milky Way (see our TGAS-APOGEE results in Table 1).  
Figure~\ref{model} shows the rotation velocity, V, and [Fe/H] of the selected particles, 
where metallicity has been assigned with the above formula. The figure includes 
the mean values and the dispersion for the binned data as a function of [Fe/H], 
and the line of best fit, corresponding to dV/d[Fe/H]$=-16.9\pm0.2$~km~s$^{-1}$~kpc$^{-1}$. 
The inferred dV/d[Fe/H] is qualitatively 
consistent with what we observed for the thin disk stars in Section~\ref{analysis}. 
Our only assumptions were negative radial and vertical metallicity gradients. We tested and found 
that assuming a zero vertical metallicity gradient does not change the slope of the derived V-[Fe/H]
 relationship.  Therefore, the negative value of dV/d[Fe/H] is mainly driven by the 
negative radial metallicity gradient. 

Because of the epicyclic motion of the stars, stars 
moving with a larger azimuthal velocity  in the Solar neighborhood tend to be 
close to their pericenter phase, meaning that their 
guiding center is larger than the Solar radius and therefore they tend to come from outer 
radii. Conversely, stars rotating slower preferentially have a guiding center 
smaller than the solar radius. Hence, if there is a negative metallicity gradient as a 
function of the guiding center or the mean radius, R$_{\rm m}$this trend can drive a 
negative slope in the V-[Fe/H] relation (see also Vera-Ciro et al. 2014). 

From this simple model, we can say that the observed negative slope in the 
V-[Fe/H] relation can be simply explained by the epicyclic motion of stars,  
given the observed radial metallicity gradient. 
Although it is often mentioned that the negative value of dV/d[Fe/H] for 
 thin disk stars is an evidence of radial migration, it can be explained solely
by the epicyclic motion (blurring), and does not require changes of 
angular momentum of the stars (churning, see Selwood \& Binney 2002).

\begin{figure}[t!]
\centering
{\includegraphics[width=8cm]{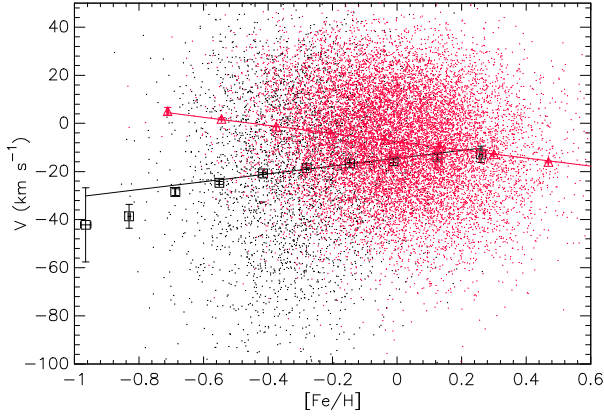}
}
\caption{Galactic rotation velocity derived for the particles in the simulations for
the thin (red) and thick disk (black). The particle's velocities are
treated similarly as the observed stars described in Section \ref{analysis}, and 
only one in ten particles are shown in the Figure. The linear fittings give 
dV/d[Fe/H] $=-16.9 \pm 0.2$ km s$^{-1}$ (thin disk) and dV/d[Fe/H]$=+16\pm 1$ km s$^{-1}$ (thick disk).}
\label{model}
\end{figure}

We then analyzed the V-[Fe/H] relation for  particles that are initially 
assigned to the thick disk population in a similar way. 
The thick disk population of the Milky Way shows a flat radial metallicity gradient, 
but a negative vertical metallicity gradient approximately 
$d{\rm [Fe/H]}/dz=-0.1$
(Mikolaitis et al. 2014; Hayden et al. 2015).

As discussed in \S~\ref{analysis}, the chemically defined thick disk 
stars have a correlation between
[Fe/H] and orbital eccentricity, $e$, with higher $e$ stars having lower [Fe/H]. Hence we assigned  metallicity to the 
thick disk particles following the relationship 
[Fe/H]$=-0.35 \times (e-0.2) -0.1 \times |z|-0.01\times R_{\rm m}-0.2$~dex 
with a dispersion of 0.175 dex. 

We found that the negative value of d[Fe/H]/d$e$ leads to a slight positive radial metallicity gradient, 
because the velocity dispersion decreases with Galactocentric distance, 
and therefore the stars with a smaller 
guiding center tend to have higher eccentricity. Hence, we applied a shallow negative radial 
metallicity gradient, to make the overall radial metallicity gradient flat as observed 
(Bensby et al. 2011; Mikolaitis et al. 2014; Hayden et al. 2015). 
We have checked that the metallicity distributions of this model match well with the metallicity 
distribution function at the different radii and vertical heights of 
[$\alpha$/Fe]$>0.17$~dex stars in the APOGEE DR12 data (Hayden et al. 2014, 2015). 

Again, we  selected the thick disk particles within a ring at 
$7.5<$R$<8.5$~kpc and $|$Z$|<0.5$~kpc.  
The velocity dispersion of the selected sample of particles is $(\sigma_{\rm R}, \sigma_{\rm \phi}, 
\sigma_{\rm z})=(40, 34, 39)$ km~s$^{-1}$. Although $\sigma_{\rm \phi}$ and $\sigma_{\rm z}$ 
are consistent with the observed velocity dispersions, the radial velocity dispersion, 
$\sigma_{\rm R}$, is significantly smaller than the observed $\sigma_{\rm R}$ for the 
thick disk population. The V-[Fe/H] relation for thick disk particles 
is shown in Fig. \ref{model}. The mean values after binning the data as 
in our analysis of the TGAS-APOGEE data are also shown, and so is the best linear fit. 
A positive slope  dV/d[Fe/H] $=16 \pm 1$ km s$^{-1}$ dex$^{-1}$
 is found for this simple model, qualitatively consistent 
with the observed positive slope for the thick disk stars in our TGAS-APOGEE sample. 
The positive dV/d[Fe/H] of this model is driven by the 
assumed eccentricity-metallicity relation, lower metallicity stars having orbits 
with a higher eccentricity and therefore a slower mean rotation velocity. This 
trend can be interpreted in a scenario in which the thick disk initially formed 
from lower metallicity and kinematically hotter gas disk, and gradually became 
more metal rich and kinematically colder, perhaps due to more gas-rich minor 
mergers at a higher redshift, as expected due to hierarchical clustering 
 (e.g., Brook et al. 2004, 2012).
  
Another way of creating a steep positive dV/d{\rm [Fe/H]} is applying a 
positive radial metallicity gradient with Rm (Curir et al. 2012). 
This would need to have the opposite slope 
to that we adopted for the thin disk particles. For the same reason given 
for the thin disk particles, a positive d[Fe/H]/dRm
can drive a positive $dV_{\phi}/d{\rm [Fe/H]}$.
However, this model disagrees with the observed flat radial metallicity gradient and 
the negative vertical metallicity gradient of the thick disk of the Milky Way.

Our simple model is sufficient to support a qualitative discussion whether or not a simple 
metallicity distribution difference can explain the observed trends 
of dV/d[Fe/H] for the thick and thin disk populations. Fine tuning of our 
model to quantitatively match the observational data, or to provide a unique solution,  
is beyond the scope of this paper. Still, this comparison highlights that the measurement 
of dV/d[Fe/H] provides strong constraints on the kinematical 
and chemical properties of the thick and thin disks.

\section{Summary}
\label{summary}

Combining astrometric information from the {\it Gaia}'s 
first data release and chemical abundances from 
the SDSS APOGEE survey, we measured the Galactic rotation velocity-[Fe/H] 
relation for the thick and thin disk stars identified on the basis of 
their [$\alpha$/Fe] abundance. 
We selected the sample of stars common to the two surveys with strict 
criteria of the relative parallax errors less than 0.3, a specific 
$\log g$ and $T_{\rm eff}$ range to select giant stars (for which the APOGEE abundances
are more reliable),  and [Fe/H]$>-1.0$ 
to minimize contamination from halo stars. We also defined the thick and thin 
disk more strictly by requiring [$\alpha$/Fe]$<0.1$~dex and $0.17<$[$\alpha$/Fe], 
respectively. We find that dV/d[Fe/H]$=-18\pm2$~km~s$^{-1}$~dex$^{-1}$ 
for thin disk stars and dV/d[Fe/H]$=+23\pm10$~km~s$^{-1}$~dex$^{-1}$ for 
 thick disk stars. 
We therefore confirm that the slope of the V-[Fe/H] relationship is different 
for the thin and thick disks. The negative dV/d[Fe/H] for thin disk stars 
is consistent with previous studies. However, our measurement of dV/d[Fe/H] 
for thick disk stars is flatter than what is claimed in  
previous studies. In addition, we find evidence that dV/d[Fe/H] depends on
[$\alpha$/Fe] for thick disk stars.

Stars with intermediate [$\alpha$/Fe] values are known to exhibit 
intermediate ages,  metallicity distributions, and kinematics. We find 
that they show an approximately
flat variation of the V-[Fe/H] relation, that is, an 
intermediate slope between the negative value for the thin disk and
the positive one for the thick disk. 

Using a simple N-body model, we demonstrate that the observed negative 
dV/d[Fe/H] for the thin disk can be explained by the observed  
negative metallicity gradient as a function of the mean orbital radius. 
The negative dV/d[Fe/H] can be explained solely
by the epicyclic motion of the stars (blurring), and it is not an 
evidence of radial migration with the change in angular momentum 
and guiding radius (churning). Our simple N-body model also 
demonstrates that the positive value of dV/d[Fe/H] for the thick disk can be naturally 
explained with the observed [Fe/H]-eccentricity correlation,
 with stars with higher eccentricity having lower [Fe/H]. 
This model provides a satisfactory explanation for the different 
signs of the slope of the V-[Fe/H] relationship for the thin and thick disks.

Our TGAS-APOGEE results indicate that the negative slope of the V-[Fe/H] relation for the thin disk
is robustly measured. However, dV/d[Fe/H] for the thick disk is sensitive to how  
the thick disk population is defined. Larger 
samples of stars with sufficient accuracy in their astrometric measurements
and associated chemistry are required 
to disentangle the correlations between kinematics and
abundances in the thick disk. This study highlights the synergy between astrometric
data from {\it Gaia} and high-resolution spectroscopy. Future Gaia data releases 
and ongoing ground-based spectroscopic surveys will further refine 
the measurements of the V-[Fe/H] relation for thick and thin disks, and 
at the same time will allow us to measure the chemodynamical signatures of the
different stellar populations, not only in the 
solar neighborhood, but also over a wide range of the Galactic disk, 
providing strong constraints on the formation 
theory of the thick and thin disks of the Milky Way. 

\begin{acknowledgements} 

We thank our colleagues, in particular Ignacio Ferreras and George Seabrook, for
enlightening discussions, an anonymous referee for suggestions, and Ron Drimmel 
for pointing out a missing important reference.
CAP is grateful for support from MINECO for this research through grant AYA2014-56359-P, 
and to the Severo Ochoa Excellence program for funding CAP's visit to MSSL in Summer 2016. 
DK and MC gratefully acknowledge the support of the UK's Science \& Technology 
Facilities Council (STFC Grant ST/K000977/1 and ST/N000811/1).  
The numerical Galaxy models for this paper were simulated on the UCL facility Grace, 
and the DiRAC Facilities (through the COSMOS and MSSL-Astro consortium) jointly 
funded by STFC and the Large Facilities Capital Fund of BIS.  We also acknowledge 
PRACE for awarding us access to their Tier-1 facilities.

This work has made use of data from the European Space Agency (ESA)
mission {\it Gaia}\footnote{http://www.cosmos.esa.int/gaia}, processed by
the {\it Gaia} Data Processing and Analysis Consortium 
(DPAC\footnote{http://www.cosmos.esa.int/web/gaia/dpac/consortium}). Funding
for the DPAC has been provided by national institutions, in particular
the institutions participating in the {\it Gaia} Multilateral Agreement.

Funding for the Sloan Digital Sky Survey IV has been provided by the
Alfred P. Sloan Foundation, the U.S. Department of Energy Office of
Science, and the Participating Institutions. SDSS acknowledges
support and resources from the Center for High-Performance Computing at
the University of Utah. The SDSS web site is www.sdss.org.

SDSS is managed by the Astrophysical Research Consortium for the 
Participating Institutions of the SDSS Collaboration including the 
Brazilian Participation Group, the Carnegie Institution for Science, 
Carnegie Mellon University, the Chilean Participation Group, the 
French Participation Group, Harvard-Smithsonian Center for Astrophysics, 
Instituto de Astrof\'{\i}sica de Canarias, The Johns Hopkins University, 
Kavli Institute for the Physics and Mathematics of the 
Universe (IPMU) / University of Tokyo, Lawrence Berkeley National Laboratory, 
Leibniz Institut f\"ur Astrophysik Potsdam (AIP), Max-Planck-Institut f\"ur 
Astronomie (MPIA Heidelberg), Max-Planck-Institut f\"ur Astrophysik (MPA Garching), 
Max-Planck-Institut f\"ur Extraterrestrische Physik (MPE), 
National Astronomical Observatory of China, New Mexico State University, 
New York University, University of Notre Dame, Observatorio Nacional / MCTI, 
The Ohio State University, Pennsylvania State University, 
Shanghai Astronomical Observatory, United Kingdom Participation Group, 
Universidad Nacional Aut\'onoma de M\'exico, University of Arizona, 
University of Colorado Boulder, University of Oxford, University of Portsmouth, 
University of Utah, University of Virginia, University of Washington, University of 
Wisconsin, Vanderbilt University, and Yale University.

\end{acknowledgements}

\end{document}